\documentclass[prd,reprint,showpacs,showkeys,superscriptaddress]{revtex4}
\usepackage{amsfonts}
\usepackage{amssymb}
\usepackage{amsmath}
\usepackage{graphicx}
\usepackage[font={footnotesize,it}]{caption}
\usepackage{latexsym}

\begin{document}

\title{Gravitational Lensing in a Model of Nonlinear Electrodynamics: The
case for electrically and magnetically charged compact objects}
\author{O. Gurtug}
\email{ozaygurtug@maltepe.edu.tr}
\affiliation{T. C. Maltepe University, Faculty of Engineering and Natural Sciences,
34857, Istanbul -Turkey}
\author{M. Mangut}
\email{mert.mangut@emu.edu.tr}
\affiliation{Department of Physics, Faculty of Arts and Sciences, Eastern Mediterranean
University, Famagusta, North Cyprus via Mersin 10, Turkey}

\begin{abstract}
This paper aims to investigate the astrophysical applicability of the electrically and magnetically charged black hole solutions obtained in a model of nonlinear electrodynamics proposed by Kruglov (Ann. Phys. Berlin 2017, 529, 170073). Theoretical calculations of the bending angles and gravitational redshifts from the theory of general relativity are studied numerically by using the stellar data of charged compact objects and a hypothetical quark star model. Calculations have revealed that although the theoretical outcomes differ from the linear Maxwell case, the plotted bending angles coincide with the linear case and it becomes hard to identify the effect of nonlinearity. However, the calculation of the redshift has shown that while the increase in the electric field leads to a decrease in the gravitational redshift,the presence of the strong magnetic field contributes to the gravitational redshift in an increasing manner.
\end{abstract}

\pacs{95.30.Sf, 98.62.Sb }
\keywords{Gravitational lensing, nonlinear electrodynamics, compact objects}
\maketitle

\section{Introduction}

Nonlinear electrodynamics (NED) has been proposed in classical general
relativity to overcome spacetime singularities that develop at the core of
charged black holes. The pioneering study in this context belongs to Born
and Infeld (BI) \cite{1}, which aimed to remove singularities at the
origin of point like charged particles. This remarkable consequence of
NED is provided by a bounded electric field at the center of point like
charges. In contrary, linear electrodynamics (LED) admits an unbounded
electric field, which leads to divergences in self-energy of point like
charges.\\

Throughout the years, the idea of BI has been developed further and now,
number of NED models became available in the literature. The common aim
of these models is to obtain electrically and magnetically charged black
hole solutions with interesting physical properties. These models can be
summarized as; the power law model with a Lagrangian $\mathcal{L=-\alpha F}%
^{k}$ in \cite{2}, the logarithmic model $\mathcal{L=-}\frac{1}{16\pi
\lambda }\ln \left( 1+\lambda \mathcal{F}\right) $ in \cite{3}, the
exponential models $\mathcal{L=\beta }^{2} \left( \exp \left( -\mathcal{F}/\mathcal{%
\beta }^{2}\right)-1 \right) $ and $\mathcal{L=-F}\exp \left( -\beta \mathcal{F}%
\right) $ in \cite{4} and \cite{5,6}, respectively, an arcsine model $%
\mathcal{L=-}\frac{1}{\beta }\sin ^{-1}\left( \beta \mathcal{F}\right) $ in
\cite{7} and a three parameter model $\mathcal{L=-F-}\frac{a\mathcal{F}}{%
1+2\beta \mathcal{F}}+\frac{\gamma }{2}\mathcal{G}$, in which $\mathcal{F}$
and $\mathcal{G}$ are the Maxwell invariants in \cite{8}.\\

In recent years, besides the resolution of curvature singularities in black
hole spacetimes, there is a growing interest in applying the results of NED to astrophysics with the purpose of  obtaining explanations to the associated observational outcomes. In refs. \cite%
{9,10,11,12,13,14}, NED has appealed to account for inflation problem.
The effect of NED on the gravitational lensing is also studied in several
works \cite{15,16,17}. Gravitational lensing is known to be one of the most
important predictions of Einstein's theory of general relativity, which
is corroborated by astronomical observations. This is a phenomena in which the
gravitational field produced by a compact object or a black hole bends the
emergent light from distance galaxies/stars while passing nearby.\\

It has been known from observational stellar data that the compact objects,
namely, Vela X-1, SAXJ1808.4-3658 and 4U1820-30, are categorized as charged
compact stars \cite{18}. The charge contents of these compact objects are
assumed to be proportional to their mass densities. Since these objects' mass densities are
enormous, they possess immense charge distributions. The
charge value at the surface of all aforementioned stars is estimated to be $\sim 10^{20}$
Coulombs \cite{19}. If this huge amount of charge is assumed to
be at rest, then the produced electric field in the surrounding region becomes
very high. In such a scenario, the bending angle of light is considered in
our recent study \cite{16} in a power law NED model together with the
cosmological constant. It has been shown that the bending angle is affected
by the value of the power parameter $k$ of the Lagrangian $%
\mathcal{L=-\alpha F}^{k}.$ \\

The charge content of charged compact objects is usually
distributed as a surface charge and the motion of the charge distribution
on the outer surface (which can be attributed as a surface current)
constitutes the main source of the magnetic field in the surrounding
geometry. Those compact objects with strong magnetic field in nearby
geometry are called magnetars. The observational data has also revealed that
the magnetars are spinning compact objects. Furthermore, the magnetic field in their
surrounding geometry is formed as a result of this spin. A recent study
\cite{19} has shown that some magnetars can have very strong magnetic fields
in nearby geometry in magnitude of $10^{18}-$ $10^{20}Gauss.$ On the other
hand, Quantum ED has proposed that the critical values of
electric and magnetic fields for pair-creations are $\sim $ $10^{18}V/m$ and
$\sim 10^{13}$ $Gauss,$ respectively. At this point, there exists an ambiguity: if there exists
such huge fields in nearby geometries, the pair-creation is inevitable
and should cause instability in these compact objects due to the
global unbalance of the forces between matter part and electrostatic part. However, observations
show these compact objects to be stable. This point needs further
investigation, and at this stage, it is worth to state that the performed numerical
analysis in this study ignores the possible pair creation. In view of this fact, it would not be wrong to speculate that
LED may not work well in places where electric and
magnetic fields are strong. Since the bounded electric and magnetic fields are the peculiar feature of NED, this motivates us to incorporate the results
of NED in astrophysical applications such as the bending of light and
gravitational redshift. For instance, in contrast to the linear Maxwell
theory in which the background magnetic field is not effective on the
gravitational redshift, it has been shown in \cite{23,24}\ that NED
contributes to the gravitational redshift.\\

In this paper, we consider a different model of NED proposed by Kruglov \cite%
{20}. According to this model, static spherically symmetric  magnetically
and electrically charged black hole solutions are given respectively by
Kruglov \cite{20} himself and Mazharimousavi -Halilsoy \cite{21}. The purpose of the
present study is to investigate whether the strong electric and magnetic fields predicted by this model of NED has an imprint on the astrophysically observable phenomena known as the bending angle of light. In this regard, the solutions
in \cite{20,21} will be incorporated. The solution presented in \cite{21},
will be used to investigate the effect of strong electric field on the path
of photon rays. The numerical analysis of possible astrophysical
applications are carried out for the charged compact stars whose physical properties are given in \cite{16,18}, by
assuming static configuration. To study the effect of strong magnetic field,
we will be considering quark stars yet to be discovered in conjunction with the
solution presented in \cite{20}.

Based on our current knowledge, isolated magnetic charges endure as hypothetical particles \cite{22} which are believed to be existing in quark stars. Consequently, the numerical analysis of the bending angle in our work is also hypothetical for now.\\

The paper is organized as follows. In Sec II, the model of NED and the
corresponding solutions for static spherically symmetric electrically and
magnetically charged solutions are presented. In Sec. III, the bending angle
of light is calculated for electrically and magnetically charged black
holes. In section IV, relevant astrophysical applications are studied
numerically for three observed electrically charged compact stars, whereas for the
magnetically charged case, the variation in the bending angle against
distance is studied hypothetically. The paper is completed with a conclusion and future research directions
in Sec. V.

\section{Electric and Magnetic Black Holes in a Model of Nonlinear
Electrodynamics}

Very recently, a new model of NED has been proposed by Kruglov \cite{20}, by
the following Lagrangian density,

\begin{equation}
\mathcal{L}\left( \mathcal{F}\right) \mathcal{=}\frac{-\mathcal{F}}{1-\sqrt{%
-\beta \mathcal{F}}},
\end{equation}%
\ in which $\mathcal{F=}\frac{1}{4}F_{\mu \nu }F^{\mu \nu }=\frac{B^{2}-E^{2}%
}{2}$ is the Maxwell invariant in terms of electric and magnetic fields $E$
and $B$, respectively. The parameter $\beta $ is a dimensionful positive
constant.  Note that linear Maxwell limit is restored when $\beta =$ $0$.
Furthermore, the electromagnetic two form is given by%
\begin{equation}
F=dA=\frac{1}{2}F_{\mu \nu }dx^{\mu }\wedge dx^{\nu }
\end{equation}%
where $A=A_{\mu }dx^{\mu }$ is the potential one form and $F_{\mu \nu
}=\partial _{\mu }A_{\nu }-\partial \nu A_{\mu }$ is the electromagnetic
field tensor.\\

The action in the Einstein-Nonlinear Maxwell theory is given by

\begin{equation}
I=\int d^{4}x\sqrt{-g}\left\{ \frac{R}{2\kappa ^{2}}+\mathcal{L}\left(
\mathcal{F}\right) \right\} ,
\end{equation}%
in which $R$ is the Ricci scalar and $\kappa ^{2}=8\pi G.$ The metric ansatz
of the static spherically symmetric spacetime is described in standard form
by%
\begin{equation}
ds^{2}=-f(r)dt^{2}+\frac{dr^{2}}{f(r)}+r^{2}\left( d\theta ^{2}+\sin
^{2}\theta d\varphi ^{2}\right) .
\end{equation}%
Variation of the action with respect to $g^{\mu \nu }$, yields the Einstein
field equations,%
\begin{equation}
R_{\mu }^{\nu }-\frac{1}{2}R\delta _{\mu }^{\nu }=\frac{\kappa ^{2}}{4\pi }\left(
\mathcal{L}\left( \mathcal{F}\right) \delta _{\mu }^{\nu }-\mathcal{L}_{%
\mathcal{F}}\left( \mathcal{F}\right) F_{\mu \lambda }F^{\nu \lambda
}\right) ,
\end{equation}%
in which $\mathcal{L}_{\mathcal{F}}\left( \mathcal{F}\right) =\frac{\partial
\mathcal{L}\left( \mathcal{F}\right) }{\partial \mathcal{F}}$ and variation
with respect to vector potential $A$, gives the Maxwell equation in a model
of NED as%
\begin{equation}
\partial _{\mu }\left( \mathcal{L}_{\mathcal{F}}\left( \mathcal{F}\right)
F^{\mu \nu }\right) =0.
\end{equation}

\subsection{Electrically charged black hole solution}

Pure electrically charged black hole solution to a model of Kruglov's NED
described above is obtained by Mazharimousavi and Halilsoy in \cite{21}. The
corresponding Einstein - Nonlinear Maxwell equations are solved by assuming $%
\mathcal{F=}$ $\frac{-E^{2}}{2}$, and the metric function is given by%
\begin{equation}
f(r)=1-\frac{2GM}{r}+\frac{2Gr^{2}}{3\alpha ^{2}}\left( 1+\frac{2q\alpha }{%
r^{2}}\right) ^{3/2}-\frac{2Gq}{\alpha }\left( 1+\frac{r^{2}}{3q\alpha }%
\right) .
\end{equation}%
in which $q$ and $M$ are charge and mass related integration constants and
for simplicity $\beta =2\alpha ^{2}$ is set. Depending on the values of
the metric parameters, the locations where $f(r)=0$ corresponds to the
horizon(s) of the charged black hole. Otherwise, the parameters that makes $%
f(r)\neq 0,$correspond to a naked singularity. The related electric field
is found to be
\begin{equation}
E(r)=\frac{1}{\alpha }\left( 1-\frac{1}{\sqrt{1+\frac{2q\alpha }{r^{2}}}}%
\right) .
\end{equation}%
Energy conditions have revealed that the electric field is bounded to
\begin{equation}
0<E<\frac{1}{\alpha },
\end{equation}%
for a physically acceptable solution. The asymptotic structure of the
solutions are given by
\begin{equation}
\lim_{r\rightarrow 0}f(r)=1-\frac{2Gq}{\alpha }-\frac{2G\widetilde{M}}{r}+%
\frac{G}{\alpha ^{2}}\sqrt{2q\alpha }r-\frac{2G}{3\alpha ^{2}}r^{2}+\mathcal{%
O}\left( r^{3}\right)
\end{equation}

\begin{equation}
\lim_{r\rightarrow \infty }f(r)=1-\frac{2GM}{r}+\frac{Gq^{2}}{r^{2}}-\frac{G\alpha q^{3}}{3r^{4}}+\mathcal{O}%
\left( r^{-6}\right)
\end{equation}%
in which $\widetilde{M}=M-\frac{2\sqrt{2}\left( q\alpha \right) ^{3/2}}{%
3\alpha ^{2}}.$ As a final remark about the obtained solution, the Taylor
series expansion of the solution (7) yields the following form:%
\begin{equation}
f(r)\simeq 1-\frac{2GM}{r}+\frac{Gq^{2}}{r^{2}}-\frac{Gq^{3}}{3r^{4}}\alpha +%
\frac{Gq^{4}}{4r^{6}}\alpha ^{2}+\mathcal{O}\left( \alpha ^{3}\right).
\end{equation}%
The obtained form in Eq.(12) indicates the deviation from the Reissner-N%
\"{o}rdstrom solution.

\subsection{Magnetically charged black hole solution}

Kruglov has found magnetically charged black hole solutions in the context
of NED via the model presented in \cite{20}. The metric function describing
magnetically charged black hole geometry is given by assuming $\mathcal{F=}$
$\frac{B^{2}}{2}=\frac{q^{2}}{2r^{4}}$. The corresponding metric function is
found to be%
\begin{equation}
f(r)=1-\frac{M}{r}-\frac{Q}{r}\tan ^{-1}\left( q_{2}r\right)
\end{equation}%
where $M=2GM_{0}$ is the mass related integration constant, $Q=2Gq_{1}$ is
the magnetic charge $q$ related constant parameter together with $q_{1}=%
\frac{q^{3/2}}{2^{3/2}\beta ^{1/4}}$ and $q_{2}=\frac{2^{1/4}}{\beta ^{1/4}%
\sqrt{q}}.$ The asymptotic structure of the solution is given by%
\begin{equation}
\lim_{r\rightarrow 0}f(r)=1-\frac{\sqrt{2}Gq}{\sqrt{\beta }}+\frac{2Gr^{2}}{%
\beta }-\frac{2^{3/2}Gr^{4}}{\beta ^{3/2}q}+\mathcal{O}\left( r^{6}\right)
\end{equation}

\begin{equation}
\lim_{r\rightarrow \infty }f(r)=1-\frac{2Gm_{M}}{r}+\frac{Gq^{2}}{r^{2}}-%
\frac{G\sqrt{\beta }q^{3}}{3\sqrt{2}r^{4}}-\frac{G\alpha q^{3}}{3r^{4}}+%
\mathcal{O}\left( r^{-6}\right)
\end{equation}%
in which $m_{M}=\frac{\pi q^{3/2}}{2^{7/4}\beta ^{1/4}}.$ The metric stated
in Eq. (13) admits black hole solutions if the constant parameters in the
metric function are chosen appropriately. Otherwise, the solution becomes
naked singular. Note that when $\beta =0,$ the solution transforms to
Reissner-N\"{o}rdstrom solution.

\section{BENDING OF LIGHT IN NONLINEAR ELECTRODYNAMICS}

\subsection{General formalism}

In the literature, there exist several methods for calculating the bending
angle of light when it passes near a massive object. However, it has been
found more convenient to use the method proposed by Rindler and Ishak (RI)
\cite{25}, especially, if one wants to see the effect of background fields on
the bending angle of light. Since in this paper we are interested in
investigating the effect of strong electric and magnetic fields on the bending
angle, we shall employ the method of RI.\\

This method originates from the fact that the inner product of two vectors
remains invariant under the rotation of coordinate systems. Therefore, the
angle between two coordinate directions $d$ and $\delta $
is given by the invariant formula (see figure 1 in \cite{16}),

\begin{equation}
\cos \left( \psi \right) =\frac{d^{i}\delta _{i}}{\sqrt{\left(
d^{i}d_{i}\right) \left( \delta ^{j}\delta _{j}\right) }}=\frac{%
g_{ij}d^{i}\delta ^{j}}{\sqrt{\left( g_{ij}d^{i}d^{j}\right) \left(
g_{kl}\delta ^{k}\delta ^{l}\right) }}.
\end{equation}%

In this formula, $g_{ij}$ stands for the metric tensor of the constant time
slice of the metric (4). A two-dimensional curved $(r,\varphi )$ space,
which is defined at the equatorial plane (when $\theta =\pi /2$ ) regarded
as the orbital plane of the light rays,
\begin{equation}
dl^{2}=\frac{dr^{2}}{f(r)}+r^{2}d\varphi ^{2}.
\end{equation}%
The constants of motion related to the null geodesics in the considered
spacetime are%
\begin{equation}
\frac{dt}{d\tau }=-\frac{E}{f(r)},\text{ \ \ \ \ \ }\frac{d\varphi }{d\tau }=%
\frac{h}{r^{2}},\text{ \ }
\end{equation}%
in which $\tau $ stands for proper time. Using these conserved quantities,
we have%
\begin{equation}
\left( \frac{dr}{d\tau }\right) ^{2}=E^{2}-\frac{h^{2}}{r^{2}}f(r),
\end{equation}%
and%
\begin{equation}
\left( \frac{dr}{d\varphi }\right) ^{2}=\frac{r^{4}}{h^{2}}\left( E^{2}-%
\frac{h^{2}}{r^{2}}f(r)\right) ,
\end{equation}%
where $E$ and $h$ represent energy and angular momentum, respectively. It
has been found convenient to introduce a new variable $u$, such that, $u=%
\frac{1}{r}.$ \ Using this transformation, Eq.(20) transforms to%
\begin{equation}
\frac{d^{2}u}{d\varphi ^{2}}=-uf(u)-\frac{u^{2}}{2}\frac{df(u)}{du}.
\end{equation}%
The solution of Eq.(21) will be used to define another equation in the
following way,%
\begin{equation}
A(r,\varphi )\equiv \frac{dr}{d\varphi }.
\end{equation}%
Now, if the direction of the orbit is denoted by $d$ and that of the
coordinate line $\varphi =$ constant $\delta ,$ we have%
\begin{eqnarray}
d &=&\left( dr,d\varphi \right) =\left( A,1\right) d\varphi \text{ \ \ \ \ \
\ \ }d\varphi <0,  \notag \\
\delta  &=&\left( \delta r,0\right) =\left( 1,0\right) \delta r.
\end{eqnarray}%
Using definitions in (16), we obtain

\begin{equation}
\tan \left( \psi \right) =\frac{\left[ g^{rr}\right] ^{1/2}r}{\left\vert
A(r,\varphi )\right\vert }.
\end{equation}%
The one-sided bending angle can therefore be defined as $\epsilon =\psi -\varphi
.$

\subsection{Bending angle in the electrically charged black hole geometry}

The metric describing the exterior geometry of electrically charged black
hole in a model of NED is given in Eq.(7). For the sake of simplicity, the
orbital equation for the light in this spacetime is written by setting $G=1$
and $\alpha =b/2$ in Eq.(7), which results in
\begin{equation}
f(r)=1-\frac{2M}{r}+\frac{8r^{2}}{3b^{2}}\left( 1+\frac{qb}{r^{2}}\right)
^{3/2}-\frac{4q}{b}\left( 1+\frac{2r^{2}}{3qb}\right)
\end{equation}%
Using Eq.(21), we get%
\begin{equation}
\frac{d^{2}u}{d\varphi ^{2}}+u=3Mu^{2}+\frac{4qu}{b}\left( 1-\sqrt{1+qbu^{2}}%
\right) .
\end{equation}%
By expanding the square root term, we obtain%
\begin{equation}
\frac{d^{2}u}{d\varphi ^{2}}+u\simeq 3Mu^{2}-2q^{2}u^{3}+q^{3}\alpha u^{4}.
\end{equation}

The homogeneous part of equation (27) has solution in harmonic form and is
given by $u=\ \frac{\sin \varphi }{R}.$ This solution corresponds to the
undeflected light in the absence of gravity, ( see figure 1 in  \cite{16}, it is displayed as a solid horizontal
line ). The next step is to substitute the first order homogeneous
solution to the right-hand side and solve for the full inhomogeneous
equation (27) which admits the approximate solution as%
\begin{equation}
u=\frac{1}{r}=\frac{\sin \varphi }{R}+\frac{M}{R^{2}}\left\{ \cos
^{2}\varphi +1\right\} +\frac{q^{2}}{4R^{3}}\left\{ 3\varphi \cos \varphi
-\sin \varphi \left[ 2-\cos ^{2}\varphi \right] \right\} +\frac{q^{3}b}{%
30R^{4}}\left\{ \cos ^{2}\varphi \left[ 6-\cos ^{2}\varphi \right]
+3\right\} .
\end{equation}%
We differentiate Eq.(28) with respect to $\varphi ,$ in accordance with
Eq.(22) to get $\ A(r,\varphi ),$
\begin{equation}
A(r,\varphi )=-\frac{r^{2}}{R}\cos \varphi -\frac{r^{2}}{R^{2}}\left\{ M\sin
2\varphi +\frac{q^{2}}{4R}\left[ 2\cos ^{2}\varphi -\sin ^{2}\varphi
-3\varphi \sin \varphi \right] +\frac{q^{3}b}{15R^{2}}\left[ \sin 2\varphi
\left( \cos ^{2}\varphi -3\right) \right] \right\} .
\end{equation}%
The constant parameter $R$ is called the impact parameter and it is related
to the physically meaningful area distance $r_{0\text{ }}$of closest
approach that occurs when $\varphi =\pi /2,$ which yields%
\begin{equation}
\frac{1}{r_{0}}=\frac{1}{R}+\frac{M}{R^{2}}-\frac{q^{2}}{2R^{3}}+\frac{q^{3}b%
}{10R^{4}}.
\end{equation}%
From this result, it is seen that the closest approach distance is affected
by the presence of the electric charge. When this result is compared with
the linear Maxwell case (Eq.(28) in \cite{16}), it is observed that in
addition to the first three terms on the right hand side, there is an additional
term which is associated with a nonlinearity parameter $b$. In the case of
positive charge, the closest approach distance decreases as long as the
nonlinearity parameter $b$ increases. In the case of negative charge, an adverse
effect is observed on the closest approach distance.\\

The one-sided bending angle $\epsilon $ of light is calculated by using
Eq.(24). The value of this angle is measured
relative to the coordinate planes where $\varphi =$ constant (see figure 1 in \cite{16}). For the small
bending angle, $\tan \psi _{0}\approx \psi _{0}.$ We then take $\varphi =0,$
for large distance away from the source. For this particular case, the
one-sided bending angle is
\begin{equation}
\epsilon =\psi _{0}\simeq \frac{2M}{R}\left\{ 1-\frac{2M^{2}}{R^{2}}+\frac{%
2M^{2}q^{2}}{R^{4}}-\frac{4M^{4}q^{3}b}{3R^{8}}\right\} +\mathcal{O}\left(
\frac{q^{8}M^{12}b ^{4}}{R^{24}}\right) .
\end{equation}%
The total bending angle is defined as the twice of this angle, namely, $%
2\psi _{0}.$

\subsection{Bending angle in the magnetically charged black hole geometry}

The metric describing the exterior geometry of a static magnetically charged
black hole is given by Kruglov \cite{20}, with his model of NED. We set $G=1$ as in the electrically charged case. Hence,
the metric becomes%
\begin{equation}
f(r)=1-\frac{M}{r}-\frac{Q}{r}\tan ^{-1}\left( q_{2}r\right)
\end{equation}%
where $M=2M_{0}$ is the mass related integration constant, $Q=2q_{1}$ is the
magnetic charge $q$ related constant parameter together with $q_{1}=\frac{%
q^{3/2}}{2^{3/2}\beta ^{1/4}}$ and $q_{2}=\frac{2^{1/4}}{\beta ^{1/4}\sqrt{q}%
}.$ Note that in the original paper of Kruglov, the mass related integration
constant $M_{0}$ is missing. The equation for the light in this spacetime is
obtained from Eq.(21) as
\begin{equation}
\frac{d^{2}u}{d\varphi ^{2}}+u=\frac{3M}{2}u^{2}+\frac{3Qu^{2}}{2}\tan
^{-1}\left( \frac{q_{2}}{u}\right) +\frac{Qu^{5}}{2u^{2}+2q_{2}^{2}}.
\end{equation}%
Since $u<<1$, Eq.(33) simplifies to%
\begin{equation}
\frac{d^{2}u}{d\varphi ^{2}}+u\simeq \frac{3M}{2}u^{2}+\frac{3Qu^{2}}{2}%
\frac{\pi }{2}+\frac{Qu^{5}}{2q_{2}^{2}}=\left( \frac{3M}{2}+\frac{3Q\pi }{4}%
\right) u^{2}+\frac{Qu^{5}}{2q_{2}^{2}}
\end{equation}

The first approximate solution, $u=\frac{\sin \varphi }{R},$ is substituted
back in Eq.(34) and its resulting solution for $u$ is obtained as%
\begin{equation}
u=\frac{1}{r}=\frac{\sin \varphi }{R}+\frac{1}{4R^{2}}\left\{ \left( 2M+Q\pi
\right) \left( \cos ^{2}\varphi +1\right) \right\} +\frac{Q}{q_{2}^{2}R^{5}}%
\left\{ \frac{\cos \varphi }{32}\left[ 3\sin \varphi \cos \varphi -5\varphi %
\right] +\frac{\sin \varphi }{48}\left[ 4-\cos ^{4}\varphi \right] \right\} ,
\end{equation}%
and the equation (22) becomes%

\begin{equation}
\begin{aligned}
  A(r,\varphi )&=-r^{2}\left\{\frac{\cos \varphi }{R}-\frac{\left( 2M+Q\pi
\right) \sin 2\varphi }{4R^{2}}+\frac{Q}{q_{2}^{2}R^{5}}\left\{ -\frac{\sin
\varphi }{32}\left[ 3\sin \varphi \cos \varphi -5\varphi \right] +\frac{\cos
\varphi }{32}\left[ 3\cos 2\varphi -5\right]\right.\right.\\
   &\left. \left. +\frac{\cos \varphi }{12}\left[
1+\cos ^{2}\varphi \sin ^{2}\varphi -\frac{\cos ^{4}\varphi }{4}\right]
\right\} \right\}.
\end{aligned}
\end{equation}

The closest approach distance $r_{0}$\  is calculated at $\varphi =\pi /2,$
which is found to be%
\begin{equation}
\frac{1}{r_{0}}=\frac{1}{R}+\frac{1}{4R^{2}}\left( 2M+Q\pi \right) +\frac{Q}{%
12q_{2}^{2}R^{5}}.
\end{equation}%
\  The one-sided bending angle is measured at $\varphi =0$ by using Eq.(24).
In doing this calculation, we first calculate $r$ using Eq.(35), and hence,
we get%
\begin{equation}
r=\frac{2R^{2}}{2M+Q\pi }.
\end{equation}%
This value is used in Eq.(22) to find $A(r,\varphi =0),$which yields $A(r,0)=%
\frac{-r^{2}}{R}.$Using these results in Eq.(24), we obtain the one sided
bending angle as
\begin{equation}
\epsilon =\psi _{0}=\frac{\left( 2M+Q\pi \right) }{2R}\left\{ 1-\frac{%
M\left( 2M+Q\pi \right) }{2R^{2}}-\frac{Q\left( 2M+Q\pi \right) }{2R^{2}}\tan
^{-1}\left( \frac{2q_{2}R^{2}}{2M+Q\pi }\right) \right\} ^{1/2}
\end{equation}

Using the property $\tan ^{-1}(\frac{1}{x})=\frac{\pi }{2}-\tan ^{-1}(x)$ in
Eq.(39) and expanding the resulting term gives%
\begin{equation}
\epsilon =\psi _{0}\simeq \frac{\left( 2M+Q\pi \right) }{2R}\left\{ 1-\frac{%
M\left( 2M+Q\pi \right) }{4R^{2}}-\frac{Q\pi \left( 2M+Q\pi \right) }{8R^{2}}%
+\frac{Q\left( 2M+Q\pi \right) ^{2}}{8q_{2}^{2}R^{4}}\right\} +\mathcal{O}%
\left( \frac{M^{2}Q^{3}}{q_{2}^{2}R^{8}}\right)
\end{equation}

\section{Applications in Astrophysics }

In this section, the possible astrophysical applications are considered. The
obtained bending angles for electrically and magnetically charged black
holes are studied numerically to display the effect of electric and magnetic
charges in the context of a model of NED. Our numerical
analysis for the electrically charged black hole is carried for three observed
charged compact stars whose properties are tabulated in  \cite{16,18}.
On the other hand, the numerical analysis for the magnetically charged black hole is carried out
for quark stars, which is yet to be discovered. The numerical values of mass, charge and radius of the magnetically
charged compact object are taken from \cite{26} in which the values are estimated through simulations. According to study \cite{26},
 a quark star having a mass of $1.15M_{\odot }$ with radius $ 7.1 km $ may have a very strong magnetic field in the order of  $10^{18}Gauss.$  It is important to note that our analysis ignores the possible pair creation for both electrically and magnetically charged compact objects. \\

The graphical analysis of the calculated bending angles are simulated by converting geometrized
units to the standard international units (S.I units). This is done by multiplying
the mass $ M $  with $Gc^{-2}$ and the charge $ q $ with $G^{1/2}c^{-2}\left( 4\pi
\varepsilon _{0}\right) ^{-1/2}.$ Here $G=6.67408\times
10^{-11}m^{3}kg^{-1}s^{-2}$ is the gravitational constant, $c=3\times
10^{8}ms^{-1}$ is the speed of light and $\varepsilon _{0}=8.85418\times
10^{-12}C^{2}N^{-1}m^{2}$ is the free space permittivity. Thus, the
one-sided bending angle is measured in $radians$. Furthermore,  we take $\varphi =0$ as the reference point at
which the one-sided bending angle is measured. This point corresponds to a
very large distance away from the source. The bending angle $\epsilon $ is
plotted against $x=R/R_{\ast }$, here $R_{\ast }$ denotes the radius of the
charged compact star.
\\

In Figure 1, the one - sided bending angle is plotted for three electrically charged compact objects, namely, Vela X-1, SAXJ1808.4-3658 and 4U1820-30.
 In order to clarify the contribution of electric charge, the graphs are generated for $ q=0$ case as well. It is very clear from the Figure 1 that the electric charge does contribute to the bending angle of light. \\
 The considered model of NED is a kind of model that the nonlinearity parameter $ \beta $ plays the role of interpolation in between the Schwarzchild (when $\beta \rightarrow \infty $) and Reissner - Nordstr\"{o}m (RN) ($\beta \rightarrow 0$) solutions. Hence, the calculated bending angle in this model must vary in between the corresponding bending angles in these solutions. Hence, the calculated bending angle in this model must not exceed the RN case. This expectation is clearly displayed in Figure 1. When the calculated bending angle Eq.(31) is plotted  with real observationally estimated data for three compact objects (mass, radius and charge values), the case for $ \beta \rightarrow 0 $ and $ \beta \neq 0 $ coincides. The conclusion from this observation is that as long as the gravitational lensing is concerned, the difference between the linear Maxwell case (RN) and the considered model of NED is almost indistinguishable.\\

 When $\beta$ increases, the bending angle decreases and for large values of $\beta$, the effect of charge diminishes and consequently, the electric field in nearby geometry becomes very weak. This result indicates that the bending angle of light lies in between the solid and dashed lines.
\begin{figure}
\begin{tabular}{cc}
 \includegraphics[width=55mm]{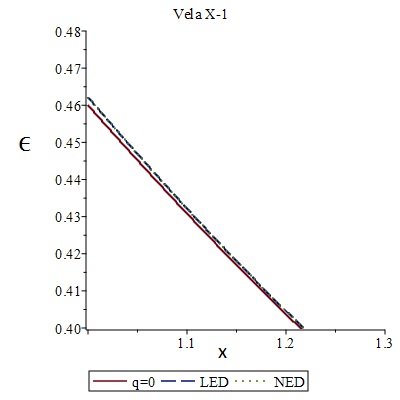}
\includegraphics[width=55mm]{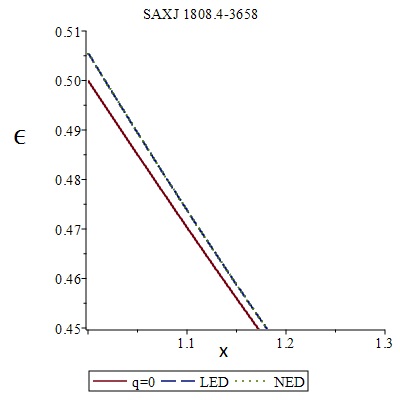}
 \includegraphics[width=55mm]{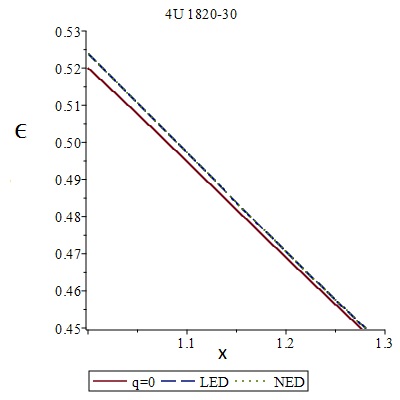}
 \end{tabular}
\centering
 \label{figure}\caption{The bending angle $\epsilon $ versus x are plotted for the electrically charged compact objects Vela X-1, SAXJ1808.4-3658, and 4U1820-30. The solid line represents the variation in the absence of charge. The coinciding dashed and the dotted lines are for LED and NED, respectively. The difference between LED and NED, as long as the bending angle of light is concerned, is almost indistinguishable. It is worth to emphasize that the graphs are generated by ignoring the possible pair creation near the surface of the objects}
\end{figure}
\newpage

Figure 2 illustrates the variation in the bending angle against a distance for a magnetically charged quark star. The graph is plotted by using Eq.(40) with the estimated numerical values of mass, radius and charge for a hypothetical quark star simulated in \cite{26}. Due to the strong magnetic field, the calculated angle is greater compared to the electrically charged counterpart.

\begin{figure}[h]
  \begin{tabular}{cc}
  \includegraphics[width=55mm]{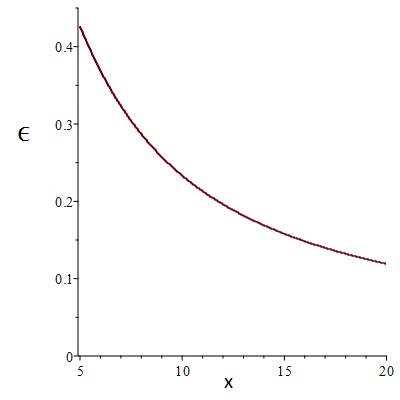}&
\end{tabular}
 \centering

  \label{figure}\caption{The bending angle $\epsilon $ versus x are plotted for a magnetically charged quark star. Graph is plotted by assuming its  mass  $1.15M_{\odot }$,  radius $ 7.1 km $ and a charge value that corresponds strong magnetic field in the order of  $10^{18}Gauss.$}

\end{figure}
\vspace{0.2cm}
Another observationally important parameter in astrophysics is the gravitational redshift produced by these electrically and magnetically charged compact objects. The exterior gravitational redshifts for these objects are calculated by taking $G=1$. Gravitational redshift for electrically charged black hole in the considered model of NED is found to be \\
\begin{equation}
  z \simeq \frac{M}{R}-\frac{R^{2}}{3\alpha^{2}}(1+\frac{2q\alpha}{R^{2}})^{3/2}+\frac{q}{\alpha}(1+\frac{R^{2}}{3q\alpha}),
\end{equation}

and for the magnetically charged black hole is calculated and given by

\begin{equation}
  z \simeq \frac{M}{R}+\frac{2q_{1}}{R}\arctan(q_{2}R),
\end{equation}

where $z = \frac{\lambda_{o}-\lambda_{e}}{\lambda_{e}}$, here $ \lambda_{o} $ denote the observed and emitted $ \lambda_{e} $ wavelengths, respectively. The gravitational redshifts are plotted for each compact object separately and displayed in Figure 3. The graphs are generated for extreme values of the nonlinearity parameter $\beta$. When $\beta$ is chosen very large, it corresponds to Schwarzchild limit such that $q=0$. When $\beta$ is very small, for instance $\beta=0$, it corresponds to the RN limit. When $\beta$ is in between these two extreme limits, the gravitational redshift lies between the solid and dashed lines. The graphs show that the gravitational redshift decreases when the intensity of the electric field increases. Of course, this is valid if the stars are in static configuration.
\begin{figure}[h]
\begin{tabular}{cc}
 \includegraphics[width=55mm]{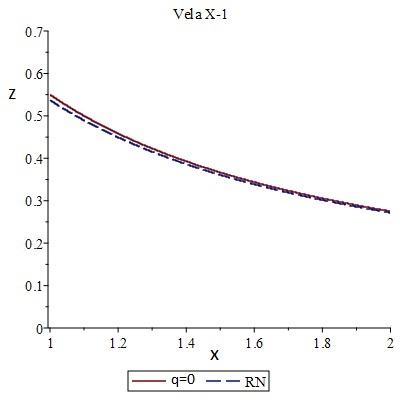}
\includegraphics[width=55mm]{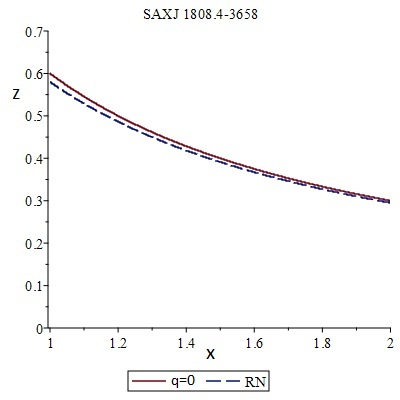}
 \includegraphics[width=55mm]{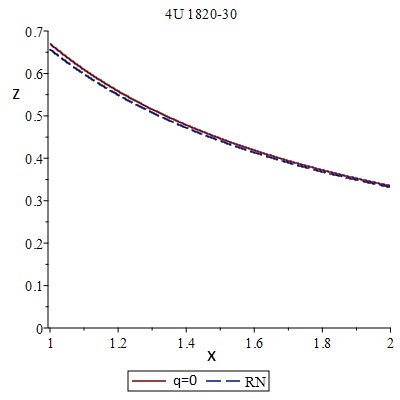}
 \end{tabular}
\centering
 \label{figure}\caption{The graphs shows the variation in the redshift against distance $x$ for each compact objects. The plots are generated for $q=0$ and $q\neq0$ cases to display the effect of electric charge. Notice that electric field decreases the gravitational redshift.}
\end{figure}
\vspace{0.2cm}

A typical neutron star (a pulsar) has no electric field, but can have a strong magnetic field. The rotating version of the solution presented in \cite{20} is not available yet, there by,  we cannot apply the data for the so far observed pulsars. The solution in \cite{20} is valid for static magnetically charged black holes or compact objects. A typical astrophysical example to this kind is quark stars that have not been observed yet. However, there are some studies in the literature where the numerical simulations reveal their estimated mass and charge values hypothetically \cite{26}. The gravitational redshift for such a quark star is studied numerically and the variation of the redshift against distance is displayed in  Figure 4. The calculated surface gravitational redshift is $ z \sim 2.4$; higher than the electrical counterpart. This result agrees with the outcomes of \cite{23}. Unlike the case in linear Maxwell theory where the background magnetic field is not effective on the gravitational redshift, we show that NED is effective and increases the gravitational redshift.
\begin{figure}[h]
 \begin{tabular}{cc}
\includegraphics[width=55mm]{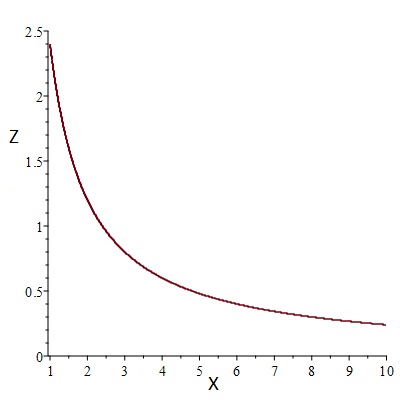}&
 \end{tabular}
 \centering
 \label{figure}\caption{Gravitational redshift is plotted against a distance $x$, for a magnetically charged quark star in a model of NED.}
\end{figure}

\section{Conclusion and Future Research Directions}

As long as the Maxwell electrodynamics is recovered in the appropriate limit, any NED theory is said to be viable. Once this perspective is persisted, the model proposed by Kruglov in \cite{20} is then viable, as the Maxwell electrodynamics is restored when the nonlinearity parameter reads $\beta \rightarrow 0$. In such cases, the following question arises naturally: ``To what extent is this model compatible with the astrophysical observational outcomes$?$" This question is the main motivation behind this article. \\

Throughout our work, the NED model proposed by Kruglov is studied in view of the applicability in astrophysical observations. In doing so, the experimentally observable quantities of astrophysics, namely, the bending angle of light and gravitational redshift are considered. The investigation involves static spherically symmetric electrically and magnetically charged black holes/compact objects in a model of Kruglov's NED. Since the solutions are static, the static configuration of electrically charged compact stars and magnetically charged quark star are taken into account. \\

The calculated bending angle of light from the solution of electrically charged black hole in the considered  model of NED is used for the charged compact objects known as Vela X-1, SAXJ1808.4-3658 and 4U1820-30. The graphs are plotted in S.I units and hence, the measured angle is expressed in radians. Figure 1 illustrates the variation in the bending angle against distance. It is clear from the graphs that the presence of electric field contributes to the bending angle of light. However, although the calculated bending angle in the considered model of NED looks different from the linear Maxwell case (i.e. RN case, Eq.(29)in \cite{16} ), once plotted with the real astrophysically observed estimated values, the two graphs are observed to overlap. Consequently, the proposed model of Kruglov's NED does not impose a significant observable change in the bending angle of light rays, as long as the gravitational lensing is of concern. \\

The variation of the bending angle against a distance for the magnetically charged black hole is generated by using the data for a quark star (yet to be discovered) where the isolated magnetic charge is believed to exist. As demonstrated in figure 2, the contribution of strong magnetic field to the bending angle is grater when compared to the electric one. \\
The gravitational redshift is another important parameter in astrophysics. The contribution of the considered model of NED is investigated for both solutions. The outcome is that the presence of electric field decreases the gravitational redshift, while the magnetic field increases. \\

In summary, we investigated the physical viability and acceptability of the model of NED proposed by Kruglov in connection with a number of charged compact star candidates and a hypothetical quark star model. The bending angle of light and the gravitational redshift are studied. The obtained results are completely plausible and agree with the earlier studies in this regard \cite{16,23,27}. However, it is very hard to claim that the considered model of NED has an observable influence on the bending angle of light. \\

As long as the light is the prime messenger from the stars and the galaxies, knowing the effect of everything else on the photon rays is extremely crucial. Thus, the research along this direction carries a vital importance. One may ask herself/himself the following question: "Is there any other mechanisms apart from the gravitational field produced by a massive astrophysical object that leads to the gravitational lensing ? For instance, to be more specific, does the light rays deviate while propagating in a universe dominated by the rotating electromagnetic fields alone or coupled with some other fields ?"  The answer is positive and our upcoming article will be on this issue. Another important point to be mentioned is the accelerated expansion of our universe and one shall recall the fact that the explanation of this phenomena with the classical theory of general relativity is not promising. Physicists are trying to solve this problem by modifying the Hilbert - Einstein gravitational action. For now, f(R), Gauss - Bonnet and the Lovelock theories remain as the most widely appealed ones among the others. The black hole solutions in theories coupled with LED or NED (for example see \cite{28}) are also important for being able to understand our universe. At this stage, another natural question arises: "What is the effect of these combined modified theories on the bending angle of light ?" In our opinion, this is an interesting problem which deserves to be investigated. There is no doubt that the bending angle equation will be much more complicated. However, the important point is to see whether these theories deviate the bending angle results obtained from the observational data recorded so far.


\begin{thebibliography}{99}
\bibitem{1} M. Born and L. Infeld, \textit{Proc. R. Soc.} \textbf{1934}, \textit{A 144}, 425.

\bibitem{2} M. Hassa\"{\i}ne and C. Mart\'{\i}nez, \textit{Class. Quant. Grav.} \textbf{2008},
\textit{25}, 195023.

\bibitem{3} H. H. Soleng, \textit{Phys. Rev\textit{. }D}  \textbf{1995},\textit{52}, 6178.


\bibitem{4} S. H. Hendi, \textit{J. High Energy Phys.} \textbf{2012}, \textit{03}, 065.

\bibitem{5} S. I. Kruglov, \textit{Ann. Phys.}  \textbf{2017},  \textit{383}, 550.

\bibitem{6} S. I. Kruglov,\textit{ EPL} \textbf{2016}, \textit{115}, 60006.

\bibitem{7} S. I. Kruglov,\textit {Ann. Phys.} \textbf{2016}, \textit{528}, 588 .

\bibitem{8} S. I. Kruglov,\textit {Phys. Lett. A}  \textbf{2015}, \textit{379}, 623.

\bibitem{9} R. Garcia-Salcedo and N. Breton, \textit{Int. J. Mod. Phys.}  \textbf{2000}, \textit{A}
\textit{15,} 4341.

\bibitem{10} C. S. Camara, M. R. de Garcia Maia, J. C. Carvalho and J. A. S.
Lima,\textit{ Phys. Rev\textit{. }D} \textbf{2004}, \textit{69}, 123504.

\bibitem{11} E. Elizalde, J. E. Lidsey, S. Nojiri and S. D. Odintsov, \textit{Phys.
Lett. B} \textbf{2003}, \textit{574}, 1.

\bibitem{12} M. Novello, S. E. P. Bergliaffa and J. M. Salim,\textit{ Phys. Rev. D} \textbf{2004}, \textit{69}, 127301.

\bibitem{13} M. Novello, E. Goulart, J. M. Salim and S. E. P. Bergliaffa,
\textit{Class. Quant. Grav.}  \textbf{2007}, \textit{24}, 3021.

\bibitem{14} D. N. Vollick, \textit{Phys. Rev. D} \textbf{2008}, \textit{78},
063524.

\bibitem{15} H. J. M. Cuesta and J. M. Salim,\textit{ Int. J. of Mod. Phys. A} \textbf{2006}, \textit{21},
43.

\bibitem{16} O. Gurtug and M. Mangut, \textit {Phys. Rev. D} \textbf{2019}, \textit{99}, 084003.

\bibitem{17} W. Javed, J. Abbas and A. Ovgun,\textit{ EPJC} \textbf{2019}, \textit{79}, 694.

\bibitem{18} M. Ilyas, \textit{Eur. Phys. J. C} \textbf{2018}, \textit{78}, 757.

\bibitem{19} S. Ray, A. L. Esp\'{\i}ndola, M. Malheiro, J. P. S. Lemos and
V. T. Zanchin,\textit{ Phys. Rev. D} \textbf{2003}, \textit{68}, 084004.

\bibitem{20} S. I. Kruglov, \textit{Ann. Phys.} \textbf{2017}, \textit{529}, 1700073.

\bibitem{21} S. H. Mazharimousavi and M. Halilsoy, \textit{Ann. Phys.} \textbf{2019},
1900236.

\bibitem{22} Y. Nambu, \textit{Phys. Rev. D} \textbf{1974},\textit{10}, 4262.

\bibitem{23} H. J. Mosquera Cuesta and J. M. Salim, \textit{Mon. Not. Roy. Astron.
Soc.} \textbf{2004}, \textit{L55}, 354.

\bibitem{24} H. J. Mosquera Cuesta and J. M. Salim, \textit{Astrophysics. J\textit{.}}  \textbf{2004},
\textit{608}, 925.

\bibitem{25} W. Rindler and M. Ishak,\textit{ Phys. Rev. D} \textbf{2007},
\textit{76}, 043006.

\bibitem{26} G. H. Bordbar and Z. Alizade,\textit{ Astrophysics} \textbf{2014}, \textit{57}, 130.
\bibitem{27} Ksh. N. Singh, N. Pant, N. Tewari and A. K. Aria,\textit{ Eur. Phys. J.} \textbf{2018}, \textit{A 54}, 77.
\bibitem{28} S. Nojiri and S. D. Odintsov, \textit{Phys. Rev. D} \textbf{2017},\textit{96}, 104008.
\end{thebibliography}
\end{document}